\documentclass[letterpaper,conference]{IEEEtran}

\usepackage[encapsulated]{CJK}
\usepackage{ucs}
\usepackage[utf8x]{inputenc}
\usepackage[cmex10]{amsmath}
\usepackage{amssymb,amscd,bbm,amsthm,mathrsfs,dsfont}
\usepackage{algorithmic,algorithm}
\usepackage{mdwmath}
\usepackage{mdwtab}
\usepackage{bm,upgreek}
\usepackage{cite}
\usepackage{graphicx,psfrag}
\usepackage{array}
\usepackage{booktabs}
\usepackage{indentfirst}
\usepackage{subfigure}
\usepackage{lipsum,fancyhdr,lastpage,refcount}
\usepackage{mathtools}
\usepackage[T1]{fontenc}
\usepackage{url}

\IEEEoverridecommandlockouts

\let\oldnl\nl
\newcommand{\nonl}{\renewcommand{\nl}{\let\nl\oldnl}}

\begin{document}

\title{Multipath Transmission Scheduling in Millimeter Wave Cloud Radio Access Networks}

\author{\IEEEauthorblockN{Xianfu Chen, Pei Liu, Hang Liu, Celimuge Wu, and Yusheng Ji}

\thanks{X. Chen is with the VTT Technical Research Centre of Finland, Finland (e-mail: xianfu.chen@vtt.fi). P. Liu is with the Department of Electrical and Computer Engineering, New York University, USA (e-mail: peiliu@nyu.edu). H. Liu is with the Department of Electrical Engineering and Computer Science, the Catholic University of America, USA (e-mail: liuh@cua.edu ). C. Wu is with the Graduate School of Informatics and Engineering, University of Electro-Communications, Tokyo, Japan (email: clmg@is.uec.ac.jp). Y. Ji is with the Information Systems Architecture Research Division, National
Institute of Informatics, Tokyo, Japan (e-mail: kei@nii.ac.jp).}

\thanks{This work was supported in part by the U.S. NSF Grant CNS-1456986 and the JSPS KAKENHI Grant JP16H02817.}
}

\maketitle

\begin{abstract}

Millimeter wave (mmWave) communications provide great potential for next-generation cellular networks to meet the demands of fast-growing mobile data traffic with plentiful spectrum available.
However, in a mmWave cellular system, the shadowing and blockage effects lead to the intermittent connectivity, and the handovers are more frequent.
This paper investigates an ``all-mmWave'' cloud radio access network (cloud-RAN), in which both the fronthaul and the radio access links operate at mmWave.
To address the intermittent transmissions, we allow the mobile users (MUs) to establish multiple connections to the central unit over the remote radio heads (RRHs).
Specifically, we propose a multipath transmission framework by leveraging the ``all-mmWave'' cloud-RAN architecture, which makes decisions of the RRH association and the packet transmission scheduling according to the time-varying network statistics, such that a MU experiences the minimum queueing delay and packet drops.
The joint RRH association and transmission scheduling problem is formulated as a Markov decision process (MDP).
Due to the problem size, a low-complexity online learning scheme is put forward, which requires no a priori statistic information of network dynamics.
Simulations show that our proposed scheme outperforms the state-of-art baselines, in terms of average queue length and average packet dropping rate.

\end{abstract}


\section{Introduction}
\label{intr}

The proliferation of wireless devices and new broadband applications has caused the demands for mobile network services to grow at an exponential rate \cite{Cisc14, Su17}.
Conventional cellular spectrum below 3 GHz is experiencing severe shortage and cannot keep up with the exponential traffic growth.
Millimeter wave (mmWave) communications, which operate at frequencies between 30 and 300 GHz, provide great potential for next-generation cellular networks to meet such demands \cite{Blei13}.
While addressing the pressing needs for additional spectrum, a mmWave cellular system raises a new set of technical challenges \cite{Rang14}.
First, the mmWave transmission characteristics, such as highly directional transmissions and low diffraction, greatly reduce the communication range and the robustness to shadowing and blockage.
As a result, the handovers are more frequent, compared with cellular systems in the legacy band \cite{Mezz16}.
These can be compensated by multi-hop relaying \cite{Zhan17} and dynamic resource allocation \cite{Qiao15}.
Second, due to high path loss, the cellular industry trend towards dense deployment makes mmWave communications more feasible and ensures network coverage \cite{Hur13}.

\begin{figure}[t]
  \centering
  \includegraphics[width=12pc]{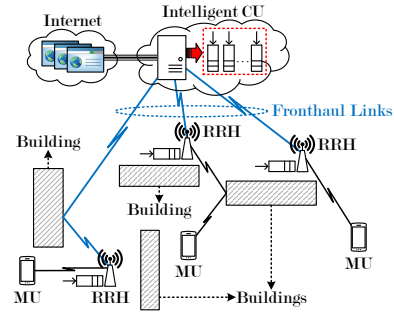}
  \caption{In an ``all-mmWave'' cloud radio access network, the fronthaul links between the intelligent central unit (CU) and the remote radio heads (RRHs) as well as the radio access links between the RRHs and the mobile users (MUs) operate over the same mmWave spectrum band.}
  \label{systMode}
\end{figure}
Recently, cloud radio access network (cloud-RAN) has been envisioned as an innovative cellular architecture to improve the network performance and reduce the cost \cite{Wu15}.
The baseband signal processing is centralized in the cloud, which allows the network operators to utilize network function virtualization techniques for resource pooling, quality-of-service (QoS) guarantee, fast handover, interference mitigation, and spectral efficiency improvement.
As illustrated in Fig. \ref{systMode}, we will exploit the potential of an ``all-mmWave" cloud-RAN architecture, where both the fronthaul \cite{Mack17} and the radio access links use mmWave frequencies.

To mitigate the impacts of intermittent mmWave links, multipath transmissions are hence orchestrated to enhance the communication reliability for mobile users (MUs).
%
%
%
A MU is able to establish multiple paths to the intelligent central unit (CU) via the remote radio heads (RRHs).
Different from prior literature (please refer to \cite{Rang14, Zhan17, Qiao15, Mezz16} and the related references therein), we propose in this paper a multipath transmission scheduling framework for an ``all-mmWave" cloud-RAN, which jointly optimizes RRH association and packet transmission scheduling. 
%
%
%
The problem of multipath transmission scheduling can be formulated as an infinite horizon Markov decision process (MDP) \cite{Rich98}, accounting for the network dynamics (i.e., the link state variations and the queue evolutions).
To address the curse of dimensionality, we decompose the original MDP into a series of MDPs with reduced state spaces and derive an on-line learning algorithm to approximate the post-decision state-value functions.
The key advantages of our proposed scheme lie in the low complexity and the no need for a priori statistics of network dynamics.

%
%
%
%
%

\section{System Model and Assumptions}
\label{sysmMode}

\begin{figure}[t]
  \centering
  \includegraphics[width=19pc]{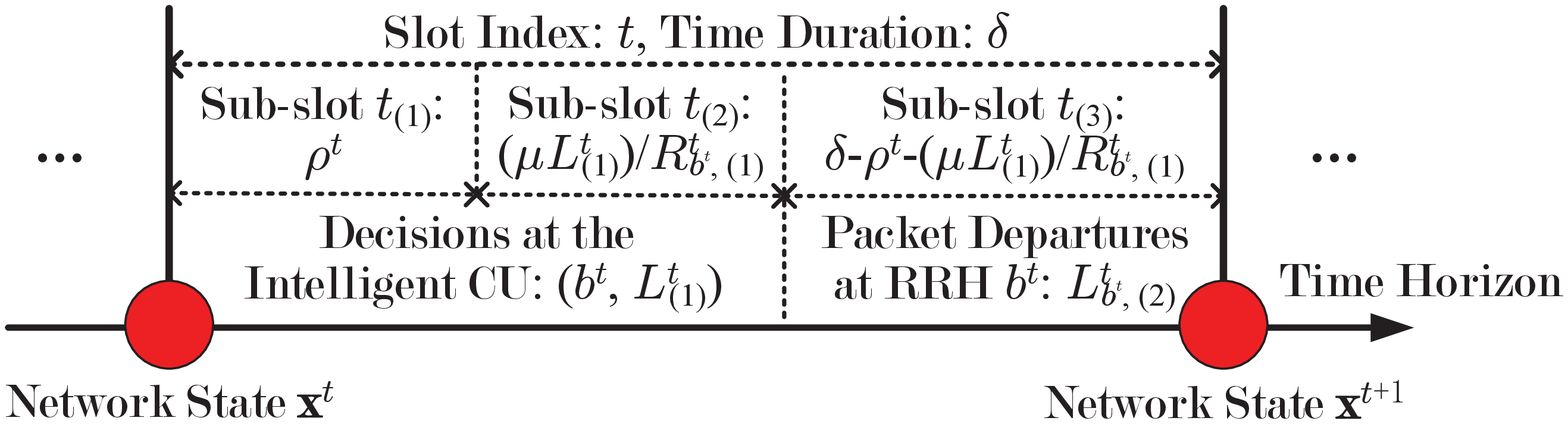}
  \caption{Structure of a scheduling slot in the multipath transmission scheduling (CU: central unit; RRH: remote radio head.).}
  \label{timeSlot}
\end{figure}

As depicted in Fig. \ref{systMode}, this paper investigates an ``all-mmWave" cloud-RAN, which is mainly composed of three components: i) an intelligent CU that strategically determines MU-RRH associations and schedules data packet transmissions, ii) the fronthaul links through which the CU communicates with the RRHs, and iii) a RAN consisting of a set $\mathcal{J} = \{1, \cdots, J\}$ of RRHs serving the MUs.
The whole system operates over a mmWave spectrum band, and we assume that during the entire communication period, a MU is exclusively allocated a portion of the bandwidth.
Hereinafter, we concentrate on a specific MU in the downlink without loss of generality.
Nevertheless, the analysis can be extended to the uplink by collecting the MU side information, such as the queue state.
The time horizon is discretized into scheduling slots, each of which is of a fixed duration $\delta$ (in seconds) and indexed by an integer $t \in \mathds{N}_+$.
Each slot $t$ can be divided into three sub-slots.
The detailed scheduling slot structure is shown in Fig. \ref{timeSlot}.
During sub-slot $t_{(1)}$, the control signalling between the CU and the MU is conducted when the RRH association changes.
The sub-slots $t_{(2)}$ and $t_{(3)}$ are for the packet transmissions between CU and RRHs and the packet transmissions between RRHs and the MU, respectively.

Let $R_{j, (1)}^t$ and $R_{j, (2)}^t$ be the state (in bits per second) of the fronthaul link between the CU and a RRH $j \in \mathcal{J}$ and that of the radio access link between the RRH $j$ and the MU during each scheduling slot $t$, which independently pick discrete values from their corresponding finite state sets $\mathcal{R}_{j, (1)}$ and $\mathcal{R}_{j, (2)}$.
At a slot $t$, the global link state $\mathbf{R}^t = \{\mathbf{R}_j^t: j \in \mathcal{J}\}$ is assumed to be perfectly known to the intelligent CU, where $\mathbf{R}_j^t = (R_{j, (1)}^t, R_{j, (2)}^t)$.
The global link state transition across the time horizon is modeled as a finite-state Markov chain with the transition probability
\begin{align}
   \textsf{Pr}\!\left\{\!\mathbf{R}^{t + 1} | \mathbf{R}^t\!\right\}\! =\!
   \prod_{j \in \mathcal{J}}\! \textsf{Pr}\!\left\{\!R_{j, (1)}^{t + 1} | R_{j, (1)}^t\!\right\}\!
   \textsf{Pr}\!\left\{\!R_{j, (2)}^{t + 1} | R_{j, (2)}^t\!\right\}\!,
\end{align}
where $\textsf{Pr}\{\Phi\}$ denotes the probability of an event $\Phi$.
%

Suppose a queue maintained at the CU and another queue at each RRH, which buffer the arriving data packets for the MU.
The packets are of equal size $\mu$ (in bits).
Let $Q_c^t$ and $Q_j^t$ denote the lengths of the queues at the CU and each RRH $j \in \mathcal{J}$ at the beginning of a scheduling slot $t$, bounded by $\overline{Q} \in \mathds{N}_+$.
At the beginning of each slot $t$, the intelligent CU first selects a RRH $b^t \in \mathcal{J}$ to serve the MU.
We assume that the MU can be associated to only one RRH during a scheduling slot.
If $b^t \neq b^{t - 1}$, handover occurs, namely, the MU is associated to the new RRH $b^t$ from the previous RRH $b^{t - 1}$.
For the two-hop link from the CU to the MU, it can be easily deduced that the time $\rho^t$ consumed by transmitting control signal during the handover procedure is of the form
\begin{align}
  \rho^t = \zeta \left(\frac{1}{R_{b^t, (1)}^t} + \frac{1}{R_{b^t, (2)}^t}\right),
\end{align}
where $\zeta \in \mathds{R}_+$ (in bits) is constant relating to the amount of control signalling data.
Otherwise, if $b^t = b^{t - 1}$, we have $\rho^t = 0$.
In other words, $\rho^t$ can be deemed as the handover cost.
The CU then determines the number $L_{(1)}^t$ of packets\footnote{Based on the global view of the link states $\mathbf{R}^t$ and the queue states $\{Q_c^t, \{Q_j^t: j \in \mathcal{J}\}\}$ at each slot $t$, the intelligent CU is able to judiciously determine $L_{(1)}^t$ and $L_{b^t, (2)}^t$ (as in (\ref{pack})) to avoid packet losses during the transmissions and packet drops at the selected RRH $b^t$.} to be transmitted to the selected RRH $b^t$, where
\begin{align}
    0 & \leq L_{(1)}^t                                                                                                                  \nonumber\\
      & \leq \min\!\left\{\!Q_c^t, \overline{Q} - Q_{b^t}^t,
        \max\!\left\{\!0, \left\lfloor\! \frac{\left(\delta - \rho^t\right) R_{b^t, (1)}^t}{\mu} \right\rfloor\!\right\}\!\!\right\},
\end{align}
with $\lfloor \cdot \rfloor$ meaning the floor function.
The RRH $b^t$ uses the time of the third sub-slot $t_{(3)}$ to deliver $L_{b^t, (2)}^t$ packets to the MU, which can be computed as
\begin{align}\label{pack}
 &   L_{b^t, (2)}^t =                                                                                                               \\
 & \min\!\!\left\{\!Q_{b^t}^t + L_{(1)}^t,
   \max\!\!\left\{\!0, \!\left\lfloor\!\frac{\Big(\delta - \rho^t - \frac{\mu L_{(1)}^t}{R_{b^t, (1)}^t}\Big) R_{b^t, (2)}^t}{\mu}\!\right\rfloor\!\right\}\!\!\right\}.                                                                               \nonumber
\end{align}

At the end of each slot $t$, $A^t \in \mathds{N}$ new packets arrive at the CU, which is assumed to be independent and identically distributed over time according to a general distribution $\textsf{Pr}\{A^t\}$.
%
%
The queue evolutions can be expressed as
\begin{align}
     Q_c^{t + 1} = \min\!\left\{\overline{Q}, Q_c^t + A^t - L_{(1)}^t\right\},
\end{align}
for the intelligent CU, and for each RRH $j \in \mathcal{J}$,
\begin{align}\label{queuDynaCent}
     Q_j^{t + 1} = Q_j^t + I_{\left\{b^t = j\right\}} \left(L_{(1)}^t - L_{j, (2)}^t\right),
\end{align}
where $I_{\{\Omega\}}$ denotes an indicator function that equals $1$ if the condition $\Omega$ is satisfied and otherwise $0$.

\section{Problem Formulation}
\label{probForm}

In this paper, we shall consider a foresighted RRH association and packet transmission scheduling problem.
For notational convenience, the global network state at each scheduling slot $t$ is represented by $\mathbf{x}^t = (Q_c^t, \mathbf{u}^t) \in \mathcal{X}$, where $\mathbf{u}^t = \{\mathbf{u}_j^t: j \in \mathcal{J}\}$ with $\mathbf{u}_j^t = (\mathbf{R}_j^t, Q_j^t)$ characterizing the local network state for the two-hop transmission link between the CU and the MU via a RRH $j$.
At the beginning of slot $t$, based on the observation of the global network state $\mathbf{x}^t$, the CU strategically decides $b^t$ and $L_{(1)}^t$ according to a stationary control policy $\bm\Theta = (\Theta_{(ra)}, \Theta_{(ts)})$, where $\Theta_{(ra)}$ and $\Theta_{(ts)}$ are, respectively, the RRH association policy and the transmission scheduling policy.
That is, $\bm\Theta(\mathbf{x}^t) = (\Theta_{(ra)}(\mathbf{x}^t), \Theta_{(ts)}(\mathbf{x}^t)) = (b^t, L_{(1)}^t)$.
Given $\bm\Theta$, the $\{\mathbf{x}^t: t \in \mathds{N}_+\}$ is a controlled Markov chain with the following state transition probability
\begin{align}\label{tran}
 & \textsf{Pr}\!\left\{\mathbf{x}^{t + 1} | \mathbf{x}^t, \bm\Theta\!\left(\mathbf{x}^t\right)\right\} =                                            \nonumber\\
 & \textsf{Pr}\!\left\{\mathbf{R}^{t + 1} | \mathbf{R}^t\right\}
   \textsf{Pr}\!\left\{\left(Q_c^{t + 1}, \mathbf{Q}^{t + 1}\right) | \left(Q_c^t, \mathbf{Q}^t\right), \bm\Theta\!\left(\mathbf{x}^t\right)\right\},
\end{align}
where $\mathbf{Q}^t = \{Q_j^t: j \in \mathcal{J}\}$.

According to the Little's law \cite{Bert87}, the average queuing delay of a stable queue is the average queue length divided by the average packet arrival rate.
For the considered single-source transmission scheduling scenario, we thus treat the average delay experienced by the MU as the average lengths of all queues at the CU and RRHs, which can be expressed as
\begin{align}
    K(\bm\Theta) =
    \lim_{\tau \rightarrow \infty} \frac{1}{\tau} \sum_{t = 1}^\tau \textsf{E}_{\bm\Theta}\!\!\left[Q\!\left(\mathbf{x}^t, b^t, L_{(1)}^t\right)\!\right],
\end{align}
where the expectation is over the randomized global network states $\mathbf{x}^t$ and the decision makings $(b^t, L_{(1)}^t)$ induced by a given control policy $\bm\Theta$, and $Q(\mathbf{x}^t, b^t, L_{(1)}^t) = \sum_{\ell \in \{c\} \cup \mathcal{J}} Q_\ell^t$. 
Moreover, due to the limited buffer size at the CU, we consider the QoS requirement from the MU as the average packet dropping rate (i.e., the long-term packet drops per scheduling slot), which is given by
\begin{align}
  & P(\bm\Theta) =                                                                                              \\
  & \lim_{\tau \rightarrow \infty} \frac{1}{\tau} \sum_{t = 1}^\tau \textsf{E}_{\bm\Theta}\!\!
    \left[\sum_{A^t} \textsf{Pr}\!\left\{A^t\right\} \max\!\left\{0, Q_c^t + A^t - L_{(1)}^t - \overline{Q}\right\}\!\right]\!.  \nonumber
\end{align}
The goal of the intelligent CU is to design an optimal control policy $\bm\Theta^*$ that minimises the average delay as well as the average packet dropping rate for the MU, which can be formally formulated as
\begin{align}\label{prob}
  \bm\Theta^* = \arg\min_{\bm\Theta} F(\bm\Theta),
\end{align}
where $F(\bm\Theta) = K(\bm\Theta) + \gamma P(\bm\Theta)$ with $\gamma \in \mathds{R}_+$ being a choice of the weight that trades off the importance of the average packet dropping rate.

\section{Solving the Optimal Control Policy}
\label{probSolv}

The formulated optimization problem in (\ref{prob}) is in general a single-agent infinite-horizon MDP with the average cost criterion.
In this section, we shall first find the optimal solution within the conventional MDP framework and then proceed to propose an approximate learning based scheme with limited network statistics information.

\subsection{Optimal MDP Solution}

Let $V_{\bm\Theta}(\mathbf{x})$ be the value function for a global network state $\mathbf{x} \in \mathcal{X}$ under a stationary control policy $\bm\Theta$.
The optimal state-value function, which is given by $V(\mathbf{x}) = V_{\bm\Theta^*}(\mathbf{x})$, $\forall \mathbf{x}$, can be achieved by solving a Bellman's optimality equation as in the following lemma \cite{Rich98}.

\noindent\textbf{Lemma 1.}
The optimal state-value function $\{V(\mathbf{x}), \forall \mathbf{x} \in \mathcal{X}\}$ satisfies the Bellman's optimality equation, that is, $\forall \mathbf{x}$,
\begin{align}\label{Bell1}
  & V(\mathbf{x}) =                                                                                                                      \\
  & \min_{b, L_{(1)}}\!\left\{f\!\left(\mathbf{x}, b, L_{(1)}\right) +
    \sum_{\mathbf{x}' \in \mathcal{X}} \textsf{Pr}\!\left\{\mathbf{x}' | \mathbf{x}, b, L_{(1)}\right\} V(\mathbf{x}')\right\} - \nu. 	 \nonumber
\end{align}
In (\ref{Bell1}), $\nu = F(\bm\Theta^*)$ and $f(\mathbf{x}, b, L_{(1)}) = Q(\mathbf{x}, b, L_{(1)}) + \gamma \sum_{A} \textsf{Pr}\{A\} \max\{0, Q_c + A - L_{(1)} - \overline{Q}\}$ is the realized cost when decisions $(b, L_{(1)})$ are performed under current global network state $\mathbf{x} = (Q_c, \mathbf{u})$, where $A$ is the number of packet arrivals, $\mathbf{u} = \{\mathbf{u}_j: j \in \mathcal{J}\}$ with $\mathbf{u}_j = (\mathbf{R}_j, Q_j)$.
$\mathbf{x}' = (Q_c', \mathbf{u}')$ represents the subsequent global network state, where $\mathbf{u}' = \{\mathbf{u}_j': j \in \mathcal{J}\}$ with $\mathbf{u}_j' = (\mathbf{R}_j', Q_j')$.

The traditional solutions to (\ref{Bell1}) are based on the value iteration or the policy iteration \cite{Rich98}.
Given the optimal function $V(\mathbf{x})$ for a global network state $\mathbf{x} \in \mathcal{X}$, we can rewrite (\ref{Bell1}) as (\ref{poli}) at the top of Page \pageref{poli},
\begin{figure*}
\begin{align}\label{poli}
 & \bm\Theta^*(\mathbf{x}) = 																							         \nonumber\\
 & \underset{b, L_{(1)}}{\arg\min}\! \left\{\sum_{j \in \mathcal{J}} I_{\{j = b\}}\!
   \sum_{\mathbf{R}'} \textsf{Pr}\!\left\{\mathbf{R}' | \mathbf{R}\right\} \sum_{A} \textsf{Pr}\{A\} \!
   \left(\!\!
   \begin{array}{c}
      \gamma \max\!\left\{0, Q_c + A - L_{(1)} - \overline{Q}\right\} - \gamma \max\!\left\{0, Q_c + A - \overline{Q}\right\} +  \\
      V\!\left(\min\!\left\{\overline{Q}, Q_c + A - L_{(1)}\right\}, \mathbf{R}',
   	  \left(Q_j + L_{(1)} - L_{j, (2)}, \mathbf{Q}_{-j}\right)\right) - 						                                 \\
   	  V\!\left(\min\!\left\{\overline{Q}, Q_c + A\right\}, \mathbf{R}', \mathbf{Q}\right)
   \end{array}\!\!
   \right)\!
   \right\}
\end{align}
\vspace{-.1cm}
\hrule
\vspace{-.5cm}
\end{figure*}
where $\mathbf{R} = \{\mathbf{R}_j: j \in \mathcal{J}\}$, $\mathbf{R}' = \{\mathbf{R}_j': j \in \mathcal{J}\}$, $\mathbf{Q} = \{Q_j: j \in \mathcal{J}\}$, $-j$ denotes all the other RRHs in set $\mathcal{J}$ except the RRH $j$, and $L_{j, (2)}$ is the number of packet departures at RRH $j$ at current slot.

\textit{Remark 1:}
The size $X$ of the global network state space $\mathcal{X}$ can be calculated as $X = (1 + \overline{Q})^{1 + J} \prod_{j \in \mathcal{J}} |\mathcal{R}_{j, (1)}| |\mathcal{R}_{j, (2)}|$, where $|\mathcal{Y}|$ means the cardinality of the set $\mathcal{Y}$.
It can be observed that $X$ grows exponentially as the number $J$ of RRHs increases.

\textit{Remark 2:}
Solving (\ref{Bell1}) not only needs \emph{complete knowledge} of the link state transition probabilities and the packet arrival statistics but suffers from \emph{exponential computation complexity} due to the extremely huge global network state space even with a reasonable number of RRHs.
%
%
%

The next subsection thereby focuses on developing a practically efficient scheme with low-complexity to achieve a near optimal control policy.

\subsection{Approximate Learning Scheme}

To tackle the first technical challenge in Remark 2, namely, the requirement of complete information of dynamic network statistics, a post-decision network state $\tilde{\mathbf{x}} \in \mathcal{X}$ as in \cite{Salo08, Chen17} is defined for each current scheduling slot.
In specific, we let $\tilde{\mathbf{x}} = (\tilde{Q}_c, \tilde{\mathbf{u}})$, where $\tilde{Q}_c = Q_c - L_{(1)}$, $\tilde{\mathbf{R}} = \mathbf{R}$, and $\tilde{Q}_j = Q_j'$, $\forall j \in \mathcal{J}$.
The optimal state-value function satisfying (\ref{Bell1}) can be hence reexpressed by: $\forall \mathbf{x} \in \mathcal{X}$,
\begin{align}\label{Bell2}
   V(\mathbf{x}) =
   \min_{b, L_{(1)}}\!\left\{Q\!\left(\mathbf{x}, b, L_{(1)}\right) + \tilde{V}\!\left(\tilde{\mathbf{x}}\right)\right\},
\end{align}
where $\tilde{V}(\tilde{\mathbf{x}})$ is termed as the optimal post-decision state-value function satisfying the Bellman's optimality equation in (\ref{postBell}) at the top of Page \pageref{postBell},
\begin{figure*}
\begin{align}\label{postBell}
   \tilde{V}(\tilde{\mathbf{x}}) =
   \sum_{\mathbf{R}'} \textsf{Pr}\{\mathbf{R}' | \mathbf{R}\} \sum_{A} \textsf{Pr}\{A\} \left(\gamma \max\!\left\{0, Q_c + A - L_{(1)} -
   \overline{Q}\right\} + V\!\left(\min\!\left\{\overline{Q}, Q_c + A - L_{(1)}\right\}, \mathbf{R}', \mathbf{Q}'\right)\right) - \nu
\end{align}
\vspace{-.1cm}
\hrule
\vspace{-.5cm}
\end{figure*}
where $\mathbf{Q}' = \{Q_j': j \in \mathcal{J}\}$.
From (\ref{Bell2}), the optimal state-value function can be directly obtained from the optimal post-decision state-value function by performing minimisation over all feasible RRH association and transmission scheduling decisions.
With (\ref{Bell2}), the calculation of optimal control policy in (\ref{poli}) can be transformed into (\ref{postpoli}), which is shown at the top of Page \pageref{postpoli}.
\begin{figure*}
\begin{align}\label{postpoli}
   \bm\Theta^*(\mathbf{x}) =
   \underset{b, L_{(1)}}{\arg\min}\!\left\{\sum_{j \in \mathcal{J}} I_{\{j = b\}} \left(
   \tilde{V}\!\left(Q_c - L_{(1)}, \mathbf{R}, \left(Q_j + L_{(1)} - L_{j, (2)}, \mathbf{Q}_{-j}\right)\right)
 - \tilde{V}(\mathbf{x})\right)\right\}
\end{align}
\vspace{-.1cm}
\hrule
\vspace{-.5cm}
\end{figure*}

From the facts underlying in (\ref{postpoli}): i) the RRH association and the transmission scheduling decisions are made sequentially but in centralized way; and ii) there exists no coupling in the packet transmissions among the RRHs, we are hence motivated to linearly decompose the optimal post-decision state-value function.
Mathematically, $\forall \tilde{\mathbf{x}} \in \mathcal{X}$,
\begin{align}\label{Deco}
   \tilde{V}\!\left(\tilde{\mathbf{x}}\right) = \tilde{V}_c\!\left(\tilde{Q}_c\right) +
   \sum_{j \in \mathcal{J}} \tilde{V}_\ell\!\left(\tilde{\mathbf{u}}_\ell\right),
\end{align}
where $\tilde{\mathbf{u}}_j = (\mathbf{R}_j, \tilde{Q}_j)$, $\forall j \in \mathcal{J}$.
Given the optimal control policy $\bm\Theta^*$, the post-decision state-value function $\tilde{V}_c(\tilde{Q}_c)$ satisfies
\begin{align}\label{postBellCU}
     \tilde{V}_c\!\left(\tilde{Q}_c\right)
 & = \sum_{A} \textsf{Pr}\{A\} \left(\gamma \max\!\left\{0, \tilde{Q}_c + A - \overline{Q}\right\} + V_c\!\left(Q_c'\right)\right) \nonumber\\
 & - \nu_c,
\end{align}
and $\forall j \in \mathcal{J}$, $\tilde{V}_j(\tilde{\mathbf{u}}_j)$ satisfies
\begin{align}\label{postBellRRH}
   \tilde{V}_j\!\left(\tilde{\mathbf{u}}_j\right) =
   \sum_{\mathbf{R}_j'} \textsf{Pr}\{\mathbf{R}_j' | \mathbf{R}_j\}V_j\!\left(\mathbf{R}_j', Q_j'\right) - \nu_j,
\end{align}
where $\nu_\ell$ ($\ell \in \{c\} \cup \mathcal{J}$) is the local optimal long-term average cost and the optimal state-value functions $V_c(Q_c')$ and $V_j(\mathbf{u}_j')$ are derived from the following
\begin{align}
   V_c\!\left(Q_c'\right)                & = Q_c + \tilde{V}_c\!\left(\tilde{Q}_c'\right),                        \label{BellCU}\\
   V_j\!\left(\mathbf{R}_j', Q_j'\right) & = Q_j + \tilde{V}_j\!\left(\mathbf{R}_j', \tilde{Q}_j'\right),         \label{BellRRH}
\end{align}
with $\tilde{Q}_c'$ and $\tilde{Q}_j'$ being the local post-decision queue states at the subsequent scheduling slot.

\emph{Remark 3:}
For the proposed linear decomposition of the post-decision state-value function, there are two main advantages.
First, in order to deploy a control policy based on the global network state $\mathbf{x} \in \mathcal{X}$, the intelligent CU has to record the state-value function with $X$ values.
Using (\ref{Deco}), only $(1 + \overline{Q}) (1 + \sum_{j \in \mathcal{J}} |\mathcal{R}_{j, (1)}| |\mathcal{R}_{j, (2)}|)$ ($\ll X$) values need to be stored, resulting in the simplified RRH association and transmission scheduling decision makings.
Second, the solving of a complex post-decision Bellman's optimality equation (\ref{postBell}) is broken into much simpler MDPs.
The linear decomposition approach is a special case of the feature-based decomposition method, but provides an accuracy guarantee of the approximation of the state-value function \cite[Theorem 2]{Tsit96}.

By replacing the post-decision state-value function in (\ref{postpoli}) with (\ref{Deco}), we arrive at a near optimal approximate control policy $\bm\Theta^*$, which includes the decisions of RRH association $\Theta_{(ra)}^*(\mathbf{x}) = b^*$ and transmission scheduling $\Theta_{(ts)}^*(\mathbf{x}) = L_{(1)}^*$ under each global network state $\mathbf{x} \in \mathcal{X}$, and can be carried out in the following two steps.

\vspace{.1cm}
\noindent\fbox{\begin{minipage}{20.4749984741210937499999999999pc}
\textbf{Step-I:} Determine the optimal number $L_{j, (1)}^*$ of packets to be scheduled to enter the queue at a RRH $j \in \mathcal{J}$ as $L_{j, (1)}^* =$ $\arg\min_{L_{(1)}} W_j(L_{(1)})$, where
\begin{align}\label{optipack}
     W_j\!\left(L_{(1)}\right)
 & = \tilde{V}_c\!\left(Q_c - L_{(1)}\right) + \tilde{V}_j\!\left(\mathbf{R}_j, Q_j + L_{(1)} - L_{j, (2)}\right) \nonumber\\
 & - \tilde{V}_c(Q_c) - \tilde{V}_j(\mathbf{u}_j).
\end{align}
\textbf{Step-II:} Select the optimal RRH $b^*$ for serving the MU to be $b^* = \arg\min_{j \in \mathcal{J}} W_j(L_{j, (1)}^*)$, then $L_{(1)}^* = L_{b^*, (1)}^*$.
\end{minipage}}


\vspace{.1cm}
As we are aware, the link states during the next scheduling slot and the number of packet arrivals at the end of current slot are unavailable beforehand.
In this case, instead of directly computing the post-decision state-value functions as in (\ref{postBellCU}) and (\ref{postBellRRH}), we propose an on-line learning algorithm to approach $\tilde{V}_c(\tilde{Q}_c)$ and $\tilde{V}_j(\tilde{\mathbf{u}}_j)$, $\forall j \in \mathcal{J}$.
Based on the observations of global network state $\mathbf{x}^t$, number of packet arrivals $A^t$ and number of packet drops $\max\{0, Q_c^t + A^t - L_{(1)}^{*, t} - \overline{Q}\}$, the decisions of RRH association $b^{*, t}$ and transmission scheduling $L_{(1)}^{*, t}$ at current scheduling slot $t$, and the resulting global network state $\mathbf{x}^{t + 1}$ at next slot $t + 1$, the intelligent CU updates the post-decision state-value functions on the fly according to (\ref{postBellCUUpda}) at the top of Page \pageref{postBellCUUpda} and
\begin{figure*}
\begin{align}\label{postBellCUUpda}
      \tilde{V}_c^{t + 1}\!\left(Q_c^t - L_{(1)}^{*, t}\right)
    = \left(1 - \alpha^t\right) \tilde{V}_c^t\!\left(Q_c^t - L_{(1)}^{*, t}\right)
    + \alpha^t \left(\gamma \max\!\left\{0, Q_c^t + A^t - L_{(1)}^{*, t} - \overline{Q}\right\} +
      V_c^t\!\left(Q_c^{t+1}\right) - \tilde{V}_c^t\!\left(\tilde{Q}_c^{(ref)}\right)\right)
\end{align}
\vspace{-.3cm}
\hrule
\vspace{-.5cm}
\end{figure*}
\begin{align}\label{postBellRRHUpda}
      \tilde{V}_j^{t + 1}\!\left(\mathbf{R}_j^t, \tilde{Q}_j^t\right)
  & = \left(1 - \alpha^t\right) \tilde{V}_j^t\!\left(\mathbf{R}_j^t, \tilde{Q}_j^t\right)          \nonumber\\
  & + \alpha^t \left(V_j^t\!\left(\mathbf{u}_j^{t+1}\right) -
      \tilde{V}_j^t\!\left(\tilde{\mathbf{u}}_j^{(ref)}\right)\right).
\end{align}
if the MU is associated with RRH $j$ during scheduling slot $t$.
In (\ref{postBellCUUpda}) and (\ref{postBellRRHUpda}), $\alpha^t \in [0, 1)$ is the learning rate, $\tilde{Q}_c^{(ref)}$ and $\tilde{\mathbf{u}}_j^{(ref)}$ are the local reference states at the CU and the RRH $j$, and the local states $Q_c^{t+1}$ and $\mathbf{u}_j^{t+1}$ at slot $t + 1$ are evaluated, respectively, by
\begin{align}\label{BellCUEval}
     V_c^t\!\left(Q_c^{t + 1}\right)
   = Q_c^{t + 1} + \tilde{V}_c^t\!\left(\tilde{Q}_c^{t + 1}\right),
\end{align}
and
\begin{align}\label{BellRRHEval}
     V_j^t\!\left(\mathbf{u}_j^{t + 1}\right)
   = Q_j^{t + 1} + \tilde{V}_j^t\!\left(\mathbf{R}_j^{t + 1}, \tilde{Q}_j^{t + 1}\right).
\end{align}
%

The online approximate learning scheme for estimating the optimal control policy is summarized in Algorithm \ref{algo}.
\begin{algorithm}[t]
    \caption{Online Approximate Learning Scheme}
    \label{algo}
    \begin{algorithmic}[1]
        \STATE \textbf{initialize} the post-decision state value functions $\tilde{V}_c^t(\tilde{Q}_c)$, $\forall \tilde{Q}_c$, and $\tilde{V}_j^t(\tilde{\mathbf{u}}_j)$, $\forall \tilde{\mathbf{u}}_j$ and $\forall j \in \mathcal{J}$, for $t = 1$.

        \REPEAT
            \STATE At the beginning of scheduling slot $t$, the intelligent CU observes the global network state $\mathbf{x}^t$, and determines the RRH association $b^{*, t}$ and the transmission scheduling $L_{(1)}^{*, t}$ according to Step-I and Step-II. \label{algoStep1}

            \STATE After transmitting packets for the MU via the selected RRH $b^{*, t}$, the CU observes the post-decision state $\tilde{\mathbf{x}}^t = (Q_c^t - L_{(1)}^{*, t}, \tilde{\mathbf{u}}^t)$, where $\tilde{\mathbf{u}}^t = \{\tilde{\mathbf{u}}_j^t: j \in \mathcal{J}\}$ with each $\tilde{\mathbf{u}}_j^t = (\mathbf{R}_j^t, Q_j^t + I_{\{b^{*, t} = j\}} (L_{(1)}^{*, t}- L_{j, (2)}^{t + 1}))$. \label{algoStep2}

            \STATE With $A^t$ new packet arrivals at the end of slot $t$, the global network state transits to $\mathbf{x}^{t + 1}$ at the slot $t + 1$.

            \STATE The CU calculates $V_c^t(Q_c^{t + 1})$ and $V_j^t(\mathbf{u}_j^{t + 1})$ according to (\ref{BellCUEval}) and (\ref{BellRRHEval}), if $b^{*, t} = j$, where $\tilde{Q}_c^{t + 1}$ and $\tilde{Q}_j^{t + 1}$ are determined following lines \ref{algoStep1} and \ref{algoStep2}.

            \STATE The CU updates the post-decision state-value functions $\tilde{V}_c^t(Q_c^t - L_{(1)}^{*, t})$ and $\tilde{V}_j^t(\tilde{\mathbf{u}}_j^t)$ according to (\ref{postBellCUUpda}) and (\ref{postBellRRHUpda}).

            \STATE The scheduling slot index is updated by $t \leftarrow t+1$.
        \UNTIL{A predefined stopping condition is satisfied.}
    \end{algorithmic}
\end{algorithm}
And the convergence property of the proposed scheme is ensured by the theorem below.

\noindent\textbf{Theorem 1}. For any initialized post-decision state-value functions $(\{\tilde{V}_c^1 (\tilde{Q}_c): \forall \tilde{Q}_c\}, \{\tilde{V}_j^1 (\tilde{\mathbf{u}}_j): \forall \tilde{\mathbf{u}}_j, \forall j \in \mathcal{J}\})$, the learning process, which is described by Algorithm \ref{algo}, converges if $\sum_{t = 1}^\infty \alpha^t = \infty$ and $\sum_{t = 1}^\infty (\alpha^t)^2 < \infty$.

\noindent\textbf{Proof}. Since the CU obtains global network information, it performs the learning rules in a centralized way.
By approximating the post-decision state-value function $\tilde{V}(\tilde{\mathbf{x}})$ with (\ref{Deco}), $\forall \tilde{\mathbf{x}} \in \mathcal{X}$, the proof proceeds similarly to the discussions in \cite{Chen17} and is thus omitted due to the page limitation.
$\hfill\Box$

\begin{figure}[t]
  \centering
  \includegraphics[width=15.9pc]{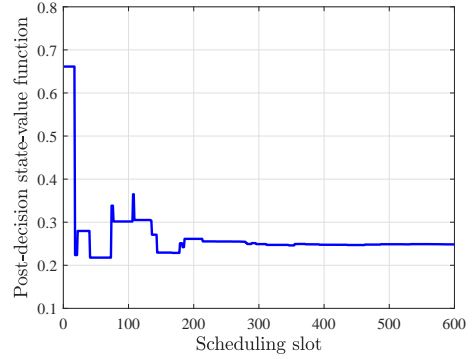}
  \caption{Convergence of a post-decision state-value function during the learning process.}
  \label{convPerf}
\end{figure}

\section{Numerical Results}
\label{simu}

This section aims to quantitatively examine the performance from our proposed scheme for multipath transmission scheduling in an ``all-mmWave" cloud-RAN.
In all simulations, the mmWave link model as in \cite{Akde14} is adopted, where three states are characterized, namely, the $\textsf{Outage}$, the line-of-sight ($\textsf{LOS}$) and the non-LOS ($\textsf{NLOS}$).
We assume that there are $J = 3$ RRHs and the values of both $R_{j, (1)}^t$ and $R_{j, (2)}^t$ for each RRH $j$ during each scheduling slot $t$ are normalized by the packet size.
In addition, $R_{j, (1)}^t$ and $R_{j, (2)}^t$ evolve according to a Markov chain model \cite{Mezz16}.
%
%
We assume that the packet arrivals to the CU queue follow a Poisson arrival process with average arrival rate $\lambda$ (in packets per scheduling slot).
For the purpose of performance comparison, the following three baseline schemes are simulated as well.
\begin{enumerate}
  \item[1)] Baseline 1: at each scheduling slot $t$, the CU selects the RRH $b^{*, t} = \arg\max_{j \in \mathcal{J}} (R_{j, (1)}^t + R_{j, (2)}^t)$ to serve the MU and schedules as many packets as possible;
  \item[2)] Baseline 2: the CU associates the MU to the RRH $b^{*, t} = \arg\max_{j \in \mathcal{J}} Q_j^t$ at the beginning of each scheduling slot $t$ for packet deliveries.
  \item[3)] Baseline 3: the CU randomly associates the MU to a RRH and randomly schedules the queued packets at the beginning of each scheduling slot $t$.
\end{enumerate}
We choose the learning rate as $\alpha(t) = \frac{\alpha_0}{\log(t) + 1}$ with $\alpha_0 = 0.6$.
Other parameters are set as: $\overline{Q} = 10$ packets, $\delta = 1$ unit time and $\zeta = 0.5$ packet .

\begin{figure}[t]
  \centering
  \includegraphics[width=15.9pc]{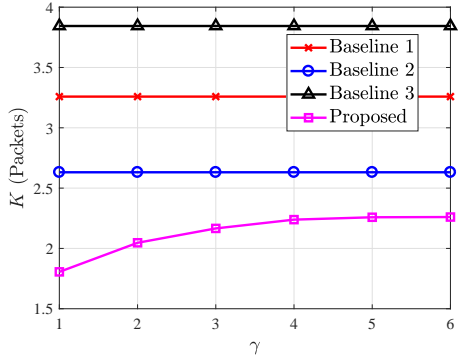}
  \caption{Average queue length versus weight $\gamma$.}
  \label{weig01}
\end{figure}

\begin{figure}[t]
  \centering
  \includegraphics[width=15.9pc]{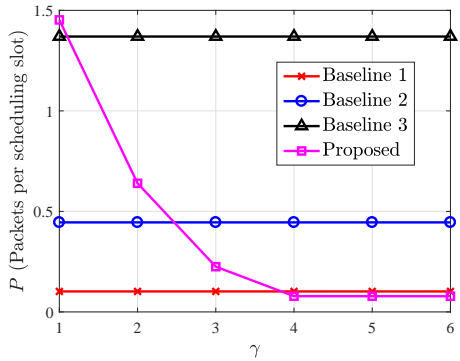}
  \caption{Average packet dropping rate versus weight $\gamma$.}
  \label{weig02}
\end{figure}

An example of the trajectory of the learning process is first plotted in Fig. \ref{convPerf} for $\tilde{V}_1^t (\textsf{LOS}, \textsf{LOS}, 3)$, from which we find that the learning process converges at a rapid speed.
Next, Figs. \ref{weig01} and \ref{weig02} plot the average queue length across all queues and the average packet dropping rate achieved from different schemes under $\lambda = 4$ and different values of $\gamma$, telling that in our proposed scheme, the average queue length increases and the average packet dropping rate decreases, as $\gamma$ increases.
A larger $\gamma$ gives higher priority to the average packet dropping rate during the learning process.
For a large enough $\gamma$, our proposed scheme has the best performance.

\begin{figure}[t]
  \centering
  \includegraphics[width=15.9pc]{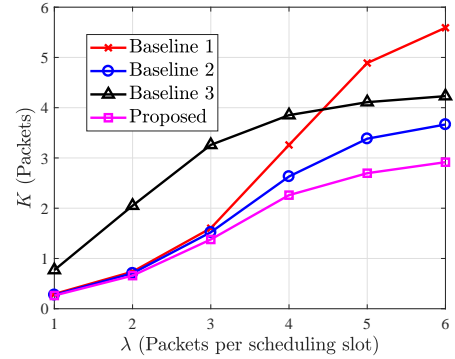}
  \caption{Average queue length versus $\lambda$.}
  \label{lamb01}
\end{figure}

\begin{figure}[t]
  \centering
  \includegraphics[width=15.9pc]{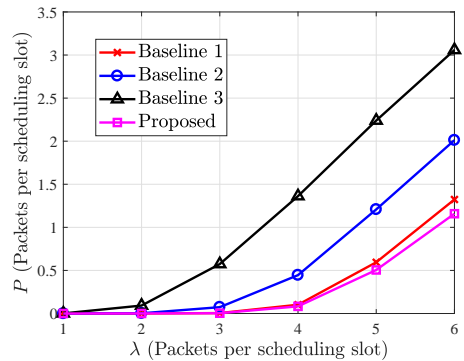}
  \caption{Average packet dropping rate versus $\lambda$.}
  \label{lamb02}
\end{figure}

In third simulation, we set $\gamma$ to be a relatively large value, i.e., $\gamma = 30$.
By increasing $\lambda$, we depict the simulated performance in Figs. \ref{lamb01} and \ref{lamb02}.
The curves exhibit that our proposed scheme outperforms the other three baselines.
The reason is that with Baselines 1--3, the CU makes shortsighted multipath transmission scheduling decisions.
Using our proposed scheme, the CU not only cares about the current transmission performance but also takes into account the performance in the future when selecting the RRH and determining the number of packets for delivery.


\section{Conclusions}
\label{conc}

In this paper, we propose a multipath transmission scheduling framework to address the technical challenges lying in unreliable links and dynamic data traffic from MUs in an ``all-mmWave" cloud-RAN.
More particularly, the problem of optimal joint RRH association and transmission scheduling for a MU is investigated and formulated as an infinite horizon MDP.
By decomposing the post-decision state-value function, we develop a low-complexity on-line learning scheme to approximate the optimal control policy.
Our proposed scheme does not need a priori knowledge of the link state transition probabilities and the data packet arrival distribution, but outperforms the baselines in literature, in terms of average queue length and average packet dropping rate.

\end{document}